\documentclass[11pt,a4paper]{article}

\usepackage{jheppub}
\usepackage{bm}

\graphicspath{{figs/}}

\title{\boldmath Non-global logarithms and fiducial transverse-momentum-dependent observables in deep-inelastic scattering}

\author[a]{Shuo Lin}
\author[a,b]{Jian Zhou}

\affiliation[a]{Key Laboratory of Particle Physics and Particle Irradiation (MOE),
Institute of Frontier and Interdisciplinary Science, Shandong University,
Qingdao, Shandong 266237, China}
\affiliation[b]{Southern Center for Nuclear-Science Theory (SCNT), Institute of
Modern Physics, Chinese Academy of Sciences, Huizhou, Guangdong 516000, China}

\emailAdd{shuolin@sdu.edu.cn}
\emailAdd{jzhou@sdu.edu.cn}

\abstract{Extracting the intrinsic transverse-momentum structure of quarks from deep-inelastic scattering data requires
  separating nonperturbative effects from perturbative radiation, which broadens the measured transverse-momentum
  distributions. We address this problem in electron--proton scattering, $ep\to eX$, by introducing the fiducial
  imbalance $\boldsymbol{q}_T$, defined as the vector sum of the scattered electron's transverse momentum and all
  hadronic transverse momenta within a specified rapidity window. This construction reduces radiative recoil without
  requiring jet reconstruction or an explicit jet veto. We account for non-global logarithms (NGLs) arising from
  correlated soft emissions across the acceptance boundary. The azimuthally averaged NGL contribution is resummed
  to all orders at leading-logarithmic accuracy in the large-$N_c$ limit using Banfi--Marchesini--Smye evolution,
  while the leading NGL correction to the first azimuthal harmonic is included at $\mathcal{O}(\alpha_s^2)$.
  For the EIC and EicC kinematics considered, widening the rapidity window increases the cross section at low $q_T$, a region particularly sensitive to nonperturbative transverse dynamics, and reduces the dilution of spin asymmetries by
  azimuthally symmetric soft recoil. Both the Sivers single-spin asymmetry $A_{UT}$ and the worm-gear double-spin
  asymmetry $A_{LT}$ increase in magnitude, while the $\boldsymbol{q}_T$ integrated cross section remains unchanged. The unpolarized $\cos\phi$ moment generated by soft gluon radiation is
  sensitive to NGL effects, with its ratio definition reducing common normalization uncertainties. The absence of
  jet reconstruction makes this observable particularly relevant at the lower collision energies of EicC,
  providing a way to study nonperturbative spin--momentum correlations by varying the fiducial acceptance.
}

\begin{document}
\maketitle
\flushbottom

\section{Introduction}
\label{sec:introduction}

Transverse-momentum-dependent (TMD) parton distributions encode correlations among partonic longitudinal momentum,
transverse momentum, and spin, providing access to the three-dimensional momentum structure of
hadrons~\cite{Angeles-Martinez:2015sea,Boussarie:2023izj}. Isolating their nonperturbative component from the
perturbative radiation that accompanies hard scattering is a central challenge in TMD physics. This separation
is especially delicate at large momentum-transfer scales, where Sudakov logarithms become large and radiative
effects can dominate the transverse-momentum spectrum even in the low-$q_T$ region most sensitive to
nonperturbative dynamics~\cite{Collins:1984kg,Boer:2001he,Boer:2017xpy}.

Two widely used probes of quark TMDs in lepton--nucleon scattering are semi-inclusive DIS with identified
final-state hadrons and the transverse-momentum imbalance between the scattered lepton and a current-region jet.
In the former, the cross section involves a convolution of a TMD distribution with a fragmentation function,
so extracting the distribution requires information about fragmentation~\cite{Bacchetta:2006tn,Ji:2004xq,Metz:2016swz}.
The electron--jet imbalance instead gives direct access to quark TMDs without fragmentation
functions~\cite{Liu:2018trl,Liu:2020dct}. Extending these measurements to polarized beams and targets provides
access to spin--orbit correlations inside the nucleon~\cite{Kang:2021ffh}. The feasibility of the electron--jet
approach has been demonstrated at HERA~\cite{H1:2021wkz,H1:2024mox}, and the forthcoming
EIC~\cite{Accardi:2012qut,AbdulKhalek:2021gbh} and EicC~\cite{Anderle:2021wcy,Xiao:2026tbs} will enable precise,
systematic measurements over a broad kinematic range.

The electron--jet imbalance does not, however, directly measure the quark's intrinsic transverse momentum:
radiation outside the jet contributes recoil that must be accounted for. The approach developed here avoids
jet reconstruction. Rather than summing the lepton and jet momenta, we sum the lepton transverse momentum and
all hadronic transverse momenta within a fixed laboratory-frame rapidity window
$\Omega=\{|\eta|<\Delta\eta/2\}$, defining the fiducial imbalance
\begin{equation*}
\boldsymbol{q}_T=\boldsymbol{l}'_T+\sum_{h\in\Omega}\boldsymbol{p}_{hT}.
\end{equation*}
By momentum
conservation, the imbalance equals the negative total transverse momentum carried by hadrons outside the
window. At leading power in the TMD regime, radiation inside the acceptance compensates the recoil of the
struck quark in the measured sum, leaving contributions from intrinsic partonic transverse momentum and
radiation outside the window. Widening the window includes more radiation in this cancellation, reducing
radiative broadening and shifting the cross section toward low $q_T$. Since the window defines the measured
momentum sum rather than an event veto, varying its width leaves the transverse-momentum-integrated cross
section unchanged for a fixed event selection.

Our work connects with recent efforts to isolate the nonperturbative content of TMDs from perturbative
radiation. Nucleon tomography through zero-jettiness measurements has been explored in
ref.~\cite{Fang:2025dee}, while restricted detector acceptance has been used to control radiation effects in
refs.~\cite{Lin:2026typ,Li:2026uug}. Reference~\cite{Lin:2026typ} examined azimuthal modulations sensitive to
gluon TMDs in tagged-jet DIS through hadronic recoil within a rapidity window; ref.~\cite{Li:2026uug} used
fiducial cuts to enhance sensitivity to gluon saturation in forward $Z^0$ production. The present work extends
these studies by incorporating non-global logarithms (NGLs), which were not included in those analyses.

NGLs arise from correlated soft emissions across the acceptance boundary. A gluon emitted inside the window
does not enter the recoil directly, but it can radiate a softer gluon outside, whose transverse momentum
then contributes to $\boldsymbol{q}_T$. Soft radiative recoil also generates azimuthal modulations, as studied
in related contexts~\cite{Hatta:2021jcd,Hatta:2020bgy,Shao:2024nor,Shao:2026doo}. At leading-logarithmic accuracy
and in the large-$N_c$ limit, successive soft branchings are resummed by the Banfi--Marchesini--Smye (BMS)
equation~\cite{Banfi:2002hw}. In soft-collinear effective theory(SCET), the corresponding resummation is organized
through the renormalization-group evolution of operators containing multiple Wilson
lines~\cite{Becher:2015hka,Becher:2016mmh}. We use BMS evolution to resum the azimuthally averaged NGL
contribution to the fiducial imbalance, while including the leading NGL correction to the $\cos\phi$ harmonic
at $\mathcal{O}(\alpha_s^2)$. We then examine how these corrections modify the azimuthal structure of the
unpolarized cross section.

As an application, we study the prospects for measuring the Sivers
function~\cite{Sivers:1989cc,Sivers:1990fh} and the worm-gear function
$g_{1T}$~\cite{Mulders:1995dh,Bacchetta:2006tn,Zhou:2009jm} using this observable. Both are central targets
of the spin programs at the EIC and EicC~\cite{Boer:2011fh,Accardi:2012qut,Anselmino:2011ay}.
The Sivers function has been constrained by global
analyses~\cite{Anselmino:2008sga,Bacchetta:2020gko,Echevarria:2020hpy,Bury:2021sue}, and extractions of
$g_{1T}$ have appeared more recently~\cite{Bhattacharya:2021twu,Yang:2024bfz}. For the kinematics considered,
we find that widening the fiducial acceptance increases both the magnitudes of the spin asymmetries and the
event yield at low $q_T$. The quark TMD channels considered here share the same spin-independent eikonal
soft factor, including its NGL contribution.
At fixed $q_T$, this soft factor does not cancel in the spin asymmetries, because the spin-dependent and
unpolarized structure functions involve different transverse-momentum convolutions.

The remainder of this paper is organized as follows. Section~\ref{sec:observables} introduces the fiducial
imbalance and discusses its kinematic properties. Section~\ref{sec:framework} develops the TMD 
framework, including Sudakov resummation and the treatment of NGLs. Section~\ref{sec:phenomenology} presents
numerical predictions for HERA, EIC, and EicC kinematics. Section~\ref{sec:conclusions} summarizes the main
findings and discusses experimental prospects. Technical details of the two-loop angular integrals and
the numerical BMS implementation are collected in appendices~\ref{app:two-emission} and~\ref{app:numerical-recipe}.

 \section{Fiducial transverse-momentum imbalance}
   \label{sec:observables}

     We consider the polarized DIS process
     \begin{equation}
     e(l,\lambda_{e})+p(p,S)\to e(l')+X,
     \end{equation}
     where $l$ and $p$ are the four-momenta of the incoming lepton and proton, $l'$ is the scattered-lepton momentum, and
     $\lambda_{e}=\pm1$ specifies the lepton helicity. The proton spin four-vector is denoted by $S$. We allow the electron beam
     to be either unpolarized or longitudinally polarized, and the proton beam to be unpolarized or transversely polarized.
     The momentum transfer is mediated by a virtual photon with spacelike momentum $q=l-l'$ and virtuality $Q^{2}=-q^{2}$.

     All kinematic quantities and fiducial criteria are evaluated in the laboratory frame, where the proton and
     electron beams collide head-on along the positive and negative $z$ axes, respectively. Neglecting beam masses,
     the center-of-mass energy squared is given by $s=(p+l)^{2}=4E_{e}E_{p}$, and the standard DIS invariants are
     \begin{equation}
     y=\frac{p\cdot q}{p\cdot l},\qquad
     x_{B}=\frac{Q^{2}}{2p\cdot q}=\frac{Q^{2}}{ys}.
     \end{equation}
     At leading order in the one-photon-exchange approximation, the hard subprocess proceeds via $\gamma^{*}q\to q$. In this
     Born configuration, the scattered lepton and the struck quark emerge almost back-to-back in the transverse plane, each carrying
     transverse momentum $P_{T}=|\boldsymbol{l}'_{T}|=Q\sqrt{1-y}$, while the pseudorapidity of the struck quark is given by
     $\eta=\ln[P_{T}/(2yE_{e})]$. In what follows, we choose the electron kinematics such that the Born-level quark is
     produced at central rapidity, $\eta=0$, which translates to the condition $P_{T}=2yE_{e}$.

     We define the fiducial imbalance by summing the transverse momenta of the scattered lepton and all hadrons within
     the fixed window $\Omega=\{|\eta|<\Delta\eta/2\}$, which covers the full azimuthal range.
     Since the beams have zero net transverse momentum, this sum is also the negative of the total transverse momentum outside the window:
     \begin{equation}
     \boldsymbol{q}_{T}
     =
     \boldsymbol{l}'_{T}
     +
     \sum_{h\in\Omega}\boldsymbol{p}_{hT}
     =-\sum_{h\notin\Omega}\boldsymbol{p}_{hT},
     \qquad
     q_{T}=|\boldsymbol{q}_{T}|,
     \label{eq:fiducial-def}
     \end{equation}

     In the TMD regime, this relates the measured imbalance to the incoming quark's
     intrinsic transverse momentum $\boldsymbol\kappa_T$:
     \begin{equation}
     \boldsymbol{q}_{T}
     =
     \boldsymbol{\kappa}_{T}
     -
     \sum_{r\notin\Omega}\boldsymbol{k}_{rT}.
     \label{eq:fiducial-decomposition}
     \end{equation}
     Here $\boldsymbol k_{rT}$ is the transverse momentum of a real QCD emission outside the rapidity window.

     Electron--jet measurements~\cite{Liu:2018trl,Liu:2020dct,Tong:2022zwp,Fang:2024auf,Tong:2023bus,H1:2021wkz}
     instead sum the transverse momenta of the scattered lepton and a reconstructed jet in the current-fragmentation region.
     The same recoil relation then holds with the radiation sum taken outside the jet rather than outside $\Omega$.
     Both imbalances therefore reduce to $\boldsymbol\kappa_T$ at Born level. The difference is the boundary:
     the jet boundary depends on the clustering algorithm, radius parameter, and dynamically determined axis, whereas the fiducial
     observable uses an inclusive sum over a fixed geometric acceptance, with no jet reconstruction.

     We choose the scattered-electron transverse momentum $\boldsymbol{l}'_{T}$ as the azimuthal reference axis; both the
     azimuthal angle $\phi$ of $\boldsymbol{q}_{T}$ and the angle $\phi_{S}$ of the proton transverse spin $\boldsymbol{S}_{T}$
     are measured relative to this direction. In electron--jet analyses, the azimuthal angle $\phi^{J}$ is
     conventionally defined with respect to the lepton--jet bisector $(\boldsymbol{l}'_{T}-\boldsymbol{p}_{JT})/2$~\cite{H1:2024mox},
     where $\boldsymbol{p}_{JT}$ is the jet transverse momentum. The two reference axes coincide in the
     back-to-back limit, with their difference suppressed by the recoil relative to the hard transverse momentum.
     We therefore use $\phi$ for both angular conventions throughout.

     We keep only the Sivers and worm-gear contributions to the spin-dependent cross section, together with the
     unpolarized terms considered here, rather than the full angular decomposition. This choice allows us to quantify
     the spin asymmetries using phenomenologically constrained distributions. Under the one-photon-exchange
     approximation, the cross section with these contributions reads~\cite{Mulders:1995dh,Bacchetta:2006tn,Kang:2021ffh,Ji:2004xq}
     \begin{equation}
     \begin{aligned}
     \frac{d\sigma}{dx_{B}\,dQ^{2}\,d^{2}\boldsymbol{q}_{T}}
     ={} & \frac{2\pi\alpha_{{\rm em}}^{2}}{x_{B}Q^{4}}
     \left[1+(1-y)^{2}\right]
     \Big[
     F_0
     +2\cos\phi\,F_1\\
      & +|\boldsymbol{S}_{T}|\sin(\phi-\phi_{S})\,F_{UT}\\
      & +\lambda_{e}|\boldsymbol{S}_{T}|
     \frac{1-(1-y)^2}{1+(1-y)^2}
     \cos(\phi-\phi_{S})\,F_{LT}
     \Big],
     \end{aligned}
     \label{eq:full-cross-section}
     \end{equation}
     The structure functions $F_0\equiv F_{UU}$ and
     $F_1\equiv F_{UU}^{\cos\phi}$ describe the azimuthal average and first harmonic of the unpolarized cross section.
     The coefficients $F_{UT}$ and $F_{LT}$ describe the single- and double-spin modulations shown above.
     The first subscript denotes the lepton polarization ($U$ for unpolarized, $L$ for longitudinally polarized),
     and $T$ denotes transverse proton polarization.
     These structure functions can be expressed in terms of quark TMDs and a soft factor accounting for radiation outside
     the acceptance window. 
We give the explicit expressions in eqs.~\eqref{eq:fw-ngl-F0}--\eqref{eq:fw-ngl-matching}, after discussing the resummation of soft-gluon radiation in section~\ref{sec:framework}.

     To isolate the various azimuthal modulations, we define the unpolarized azimuthal moment with the single- and
     double-spin asymmetries:
     \begin{align}
     \langle\cos\phi\rangle & =\frac{F_1}{F_0},
     \label{eq:cos-moment-def}\\[4pt]
     A_{UT} & =\frac{F_{UT}}{F_0},
     \label{eq:aut-sivers-def}\\[4pt]
     A_{LT} & =\frac{1-(1-y)^2}{1+(1-y)^2}\,\frac{F_{LT}}{F_0}.
     \label{eq:alt-wormgear-def}
     \end{align}
     The single-spin asymmetry $A_{UT}$ is sensitive to the Sivers distribution
     $f_{1T}^{\perp}$~\cite{Sivers:1989cc,Sivers:1990fh,Echevarria:2020hpy,Boer:2003cm,Collins:2002kn}, whereas the
     double-spin asymmetry $A_{LT}$ probes the worm-gear distribution
     $g_{1T}$~\cite{Bhattacharya:2021twu,Mulders:1995dh,Kotzinian:2006dw}. Measuring how these asymmetries evolve with the
     fiducial acceptance window offers a novel, jet-free pathway to explore the transverse-spin structure of the nucleon.

\section{Soft-gluon recoil and TMD structure functions}
\label{sec:framework}
\begingroup
% Slightly more space around displays in this equation-heavy section.
\addtolength{\abovedisplayskip}{2pt}
\addtolength{\belowdisplayskip}{2pt}
\setlength{\abovedisplayshortskip}{5pt plus 2pt minus 1pt}
\setlength{\belowdisplayshortskip}{7pt plus 2pt minus 2pt}
\setlength{\jot}{4pt}

The azimuthal structure of the fiducial transverse-momentum imbalance encodes information about both the intrinsic
  quark transverse-momentum distributions and the soft-gluon recoil that depends on the fiducial acceptance. In this section, we develop the theoretical framework for computing this type of observable. Our treatment includes Sudakov
  resummation to capture the large global logarithmic corrections, as well as the non-global logarithms that emerge when correlated gluon emissions straddle
  the acceptance boundary.
\subsection{Quark TMDs}
\label{sec:fw-tmd}

The unpolarized, Sivers, and worm-gear quark TMDs are defined through the
following leading-twist projections of the quark correlator~\cite{Mulders:1995dh,Bacchetta:2006tn}:
  \begin{align}
  \Phi^{[\gamma^{+}]}(x,\boldsymbol{k}_{T})
  &=f_{1}(x,k_{T}^{2})-\frac{\epsilon_{T}^{ij}k_{Ti}S_{Tj}}{M_{p}}f_{1T}^{\perp}(x,k_{T}^{2}),\\[3pt]
  \Phi^{[\gamma^{+}\gamma_{5}]}(x,\boldsymbol{k}_{T})
  &=\frac{\boldsymbol{k}_{T}\!\cdot\!\boldsymbol{S}_{T}}{M_{p}}g_{1T}(x,k_{T}^{2}),
  \label{eq:fw-spin-correlators}
  \end{align}
  Here $M_p$ is the proton mass and $\epsilon_T^{ij}$ is the transverse antisymmetric tensor.

  In the region $q_T\ll Q$, logarithms of $Q/q_T$ require resummation.
  Impact-parameter space provides a natural implementation for this resummation as the transverse-momentum
  convolution of the quark TMDs with soft recoil becomes a product.
  We introduce the Fourier-conjugate variable $\boldsymbol b$ to $\boldsymbol q_T$,
  with $b=|\boldsymbol b|$ and $\phi_b=\arg\boldsymbol b-\arg\boldsymbol l'_{T}$.
  The unpolarized distribution in impact-parameter space is defined by
  \begin{equation}
  \begin{aligned}
  \widetilde{f}_1(x,b)
  &\equiv\int d^{2}\boldsymbol{k}_{T}\,e^{-i\boldsymbol{b}\cdot\boldsymbol{k}_{T}}f_1(x,k_{T}^{2})
  =2\pi\int_{0}^{\infty}dk_{T}\,k_{T}\,J_{0}(bk_{T})\,f_1(x,k_{T}^{2}).
  \end{aligned}
  \label{eq:fw-tmd-fourier}
  \end{equation}

  The Sivers and worm-gear functions both correlate quark transverse momentum
  with the proton's transverse spin, but the Sivers function is naive-$T$-odd
  whereas $g_{1T}$ is naive-$T$-even. At leading power, soft radiation is
  insensitive to this distinction, so the two distributions share the same
  spin-independent TMD evolution kernel. In impact-parameter space, this
  evolution acts multiplicatively on the scalar coefficients of their distinct
  spin-dependent tensor structures. The transverse-momentum factors in both
  terms lead to $J_1$-weighted integrals for these coefficients, which we define as
  \begin{equation}
  \begin{aligned}
  \widetilde{f}_{1T}^{\perp(1)}(x,b)
  &\equiv\frac{2\pi}{M_{p}^{2}}\int_{0}^{\infty}dk_{T}\,
  \frac{k_{T}^{2}}{b}J_{1}(bk_{T})\,f_{1T}^{\perp}(x,k_{T}^{2}),\\[3pt]
  \widetilde{g}_{1T}^{(1)}(x,b)
  &\equiv\frac{2\pi}{M_{p}^{2}}\int_{0}^{\infty}dk_{T}\,
  \frac{k_{T}^{2}}{b}J_{1}(bk_{T})\,g_{1T}(x,k_{T}^{2}).
  \end{aligned}
  \label{eq:fw-first-moments}
  \end{equation}
  With the common arguments $(x,b)$ suppressed, the correlator projections take the form
  \begin{align}
\widetilde{\Phi}^{[\gamma^{+}]}
  &=\widetilde{f}_{1}+iM_{p}\epsilon_{T}^{ij}b_{i}S_{Tj}\widetilde{f}_{1T}^{\perp(1)},\\[3pt]
  \widetilde{\Phi}^{[\gamma^{+}\gamma_{5}]}
  &=-iM_{p}\,\boldsymbol{b}\!\cdot\!\boldsymbol{S}_{T}\,\widetilde{g}_{1T}^{(1)}.
    \label{eq:fw-bspace-correlators}
  \end{align}

  While the Sudakov factor resums transverse-momentum logarithms,
  the collinear scale dependence is carried by the twist-three
  correlation functions onto which spin-dependent TMDs are
  matched. To connect the Sivers and worm-gear distributions to
  these collinear functions, we
  introduce the first transverse moments:
  \begin{equation}
  f_{1T}^{\perp(1)}(x)
  =
  \int d^{2}\boldsymbol{k}_{T}\,
  \frac{k_{T}^{2}}{2M_{p}^{2}}\,
  f_{1T}^{\perp}(x,k_{T}^{2}),
  \qquad
  g_{1T}^{(1)}(x)
  =
  \int d^{2}\boldsymbol{k}_{T}\,
  \frac{k_{T}^{2}}{2M_{p}^{2}}\,
  g_{1T}(x,k_{T}^{2}).
  \label{eq:fw-transverse-moments}
  \end{equation}
  The Sivers moment is related to the diagonal Qiu--Sterman function
  $T_F(x,x)$~\cite{Qiu:1991pp,Qiu:1998ia,Zhou:2008mz}. The worm-gear moment
  equals the chiral-even twist-three function $\tilde g(x)$ in the convention
  of ref.~\cite{Zhou:2009jm}, with $g_{1T}^{(1)}(x)=\tilde g(x)$.

  The fiducial definition alters how soft radiation enters the Sudakov factor.
  When a soft gluon is emitted inside the acceptance, its transverse momentum
  contributes to the measured hadronic sum and compensates the parton recoil,
  leaving the fiducial imbalance unchanged at leading power. Radiation outside
  the acceptance, however, is omitted from the sum and produces a net measured recoil.

  The acceptance boundary thus determines which real emissions contribute to
  the global recoil radiator: inside the window, the inclusive real--virtual
  cancellation removes their contribution, while outside, the recoil measurement
  leaves a nontrivial contribution. The standard Sudakov factor must therefore
  be modified to account for this restricted angular region. In the next subsection,
  we derive the acceptance-dependent radiator and include the angular dependence
  of the soft recoil needed for the azimuthal structure functions.

  \subsection{Sudakov resummation}
\label{sec:fw-global}

  The acceptance dependence of the Sudakov factor follows from the angular phase space
  for soft recoil. At leading power, the incoming and outgoing quarks act as eikonal
  color sources, forming a single radiating dipole. The fiducial window enters by restricting the phase space of emissions that contribute to the recoil.
  We extract the resulting logarithms from the one-loop real--virtual contribution,
  with the recoil restricted to gluons outside the window.

  The incoming quark momentum is $p_{B}=x_{B}p$, and the outgoing struck-quark momentum is $p_{J}=x_{B}p+q$.
  The soft gluon carries lightlike momentum
  \begin{equation}
  k^{\mu}=(k_{T}\cosh\eta,\,k_{T}\cos\varphi,\,k_{T}\sin\varphi,\,k_{T}\sinh\eta),
  \end{equation}
  Here $\varphi$ is measured from the Born-level struck-quark direction, which serves
  as the reference axis for the angular integrals. The eikonal antenna factor is
  \begin{equation}
  S_{g}(p_{B},p_{J};k)
  \equiv\frac{2p_{B}\cdot p_{J}}{(p_{B}\cdot k)(p_{J}\cdot k)}
  =\frac{1}{k_{T}^{2}}\frac{2e^{\eta}}{\cosh\eta-\cos\varphi}.
  \label{eq:fw-eikonal-antenna}
  \end{equation}

  Only emissions with $\eta<-\Delta\eta/2$ or $\eta>\Delta\eta/2$ contribute to the
  recoil in eq.~\eqref{eq:fiducial-decomposition}. At fixed $k_T$, longitudinal
  momentum conservation imposes $k_Te^\eta<Q^2/P_T$, setting the forward rapidity
  bound $\eta<\ln[Q^2/(P_Tk_T)]$. For $k_T<Q^2e^{-\Delta\eta/2}/P_T$, this bound
  lies beyond the window. The antenna is exponentially suppressed towards negative
  rapidity, so we extend the backward integration to $-\infty$. The radiation
  contributing to the recoil occupies
  \begin{equation}
  -\infty<\eta<-\frac{\Delta\eta}{2},
  \qquad
  \frac{\Delta\eta}{2}<\eta<
  \ln\frac{Q^{2}}{P_{T}k_{T}}.
  \label{eq:fw-rapidity-support}
  \end{equation}

  The azimuthal average governs radiative broadening, while the first harmonic
  generates the unpolarized $\cos\phi$ modulation. The corresponding antenna
  integrals are
  \begin{align}
  \int_{0}^{2\pi}\frac{d\varphi}{2\pi}
  \frac{1}{\cosh\eta-\cos\varphi}
  &=
  \frac{1}{|\sinh\eta|},
  \label{eq:fw-antenna-moment-0}
  \\
  \int_{0}^{2\pi}\frac{d\varphi}{2\pi}
  \frac{\cos\varphi}{\cosh\eta-\cos\varphi}
  &=
  \frac{e^{-|\eta|}}{|\sinh\eta|}.
  \label{eq:fw-antenna-moment-1}
  \end{align}

  Using eq.~\eqref{eq:fw-antenna-moment-0} and integrating over
  the out-of-window rapidity regions in
  eq.~\eqref{eq:fw-rapidity-support}, we obtain
  \begin{align}
  &\int_{-\infty}^{-\Delta\eta/2}\!d\eta\,
  \frac{2C_{F}e^{2\eta}}{1-e^{2\eta}}
  +\int_{\Delta\eta/2}^{\ln[Q^{2}/(P_{T}k_{T})]}\!d\eta\,
  \frac{2C_{F}}{1-e^{-2\eta}}
  \nonumber\\
  &\hspace{1.5cm}
  =C_{F}\left[
  \ln\frac{Q^{2}}{k_{T}^{2}}
  +\ln\frac{Q^{2}}{P_{T}^{2}}
  +B_{0}(\Delta\eta)
  \right]
  +\mathcal{O}\!\left(\frac{P_{T}^{2}k_{T}^{2}}{Q^{4}}\right),
  \label{eq:fw-B0-integral}
  \end{align}
  where $C_F=(N_c^2-1)/(2N_c)$. The finite acceptance-dependent
  coefficient is
  \begin{equation}
  B_{0}(\Delta\eta)
  =-\Delta\eta-2\ln\!\left(1-e^{-\Delta\eta}\right).
  \label{eq:fw-B0}
  \end{equation}
  For a large rapidity interval, $B_0(\Delta\eta)\approx-\Delta\eta$.

  The $\cos\varphi$ projection in eq.~\eqref{eq:fw-antenna-moment-1} decreases
  exponentially at large positive rapidity. Extending its upper limit to $+\infty$
  introduces no logarithmic corrections and gives the finite coefficient
  \begin{equation}
  B_1(\Delta\eta)
  =-\int_{-\infty}^{-\Delta\eta/2}\!d\eta\,
  e^{2\eta}\operatorname{csch}\eta
  +\int_{\Delta\eta/2}^{\infty}\!d\eta\,
  \operatorname{csch}\eta
  =4\operatorname{atanh}\!\left(e^{-\Delta\eta/2}\right)
  -2e^{-\Delta\eta/2}.
  \label{eq:fw-B1}
  \end{equation}
  The positive coefficient $B_1$ decreases as the window widens:
  more of the forward radiation is included in the hadronic sum, reducing the
  out-of-acceptance recoil.

  The acceptance modifies both the azimuthal average through $B_0$ and the first harmonic through $B_1$.
  For the single-gluon contribution at $\boldsymbol\kappa_T=0$, $\boldsymbol q_T=-\boldsymbol k_T$;
  the opposite electron and Born-quark reference directions then give $\cos\varphi=\cos\phi$.
  The one-loop soft contribution, normalized to the Born cross section $\sigma_0$, is
  \begin{equation}
  \frac{1}{\sigma_0}
  \frac{d\sigma}{d^{2}\boldsymbol{q}_{T}}
  =
  \frac{\alpha_{s}C_{F}}{2\pi^{2}q_{T}^{2}}
  \left[
  \ln\frac{Q^{2}}{q_{T}^{2}}
  +\ln\frac{Q^{2}}{P_{T}^{2}}
  +B_{0}(\Delta\eta)
  +2B_{1}(\Delta\eta)\cos\phi
  \right].
  \label{eq:fw-one-loop-fiducial}
  \end{equation}

  The same soft-radiation mechanism governs electron--jet correlations, with the jet boundary
  replacing the acceptance boundary~\cite{Hatta:2021jcd}:
  \begin{equation}
  \frac{1}{\sigma_0}
  \frac{d\sigma^{J}}{d^{2}\boldsymbol{q}_{T}}
  =
  \frac{\alpha_{s}C_{F}}{2\pi^{2}q_{T}^{2}}
  \left[
  \ln\frac{Q^{2}}{q_{T}^{2}}
  +\ln\frac{Q^{2}}{P_{T}^{2}}
  +c_{0}(R)
  +2c_{1}(R)\cos\phi
  \right],
  \label{eq:fw-one-loop-eJ}
  \end{equation}
  For the inclusive-$k_T$ algorithm with jet radius $R=1$, $c_{0}(1)=-0.253$ and $c_{1}(1)=0.782$.
  The fiducial observable replaces these jet-radius-dependent coefficients by $B_0(\Delta\eta)$
  and $B_1(\Delta\eta)$: the acceptance window controls the soft recoil without jet reconstruction.

  The one-loop spectrum contains a logarithmically enhanced azimuthal
  average and a finite first harmonic. To resum these logarithms,
  we work in impact-parameter space, where the transverse-momentum
  convolution becomes a product and independent soft emissions
  exponentiate. TMD evolution then gives the Sudakov
  factor~\cite{Liu:2018trl,Liu:2020dct,Shao:2023zge}, while the first
  harmonic is included as a one-loop matching correction.
  With $\mu_b=b_0/b$ and $b_0=2e^{-\gamma_E}$,
  the soft integral up to $Q$ takes the following large-$b$ form:
  \begin{align}
  &C_Fg_s^2\int d^2\boldsymbol q_T
  \int_{\bar\Omega} \frac{d^3\boldsymbol k}{(2\pi)^3\,2k^0}\,
  \delta^{(2)}(\boldsymbol q_T+\boldsymbol k_T)\,
  S_g(p_B,p_J;k)
  \left[1-e^{-i\boldsymbol{b}\cdot\boldsymbol{q}_{T}}\right]
  \nonumber\\
  &\simeq
  \frac{\alpha_{s}C_{F}}{2\pi}
  \left[
  \frac{1}{2}\ln^{2}\frac{Q^{2}}{\mu_{b}^{2}}
  +
  \left(
  \ln\frac{Q^{2}}{P_{T}^{2}}
  +B_{0}(\Delta\eta)
  \right)
  \ln\frac{Q^{2}}{\mu_{b}^{2}}
  +4iB_1(\Delta\eta)\cos\phi_b+\cdots
  \right],
  \label{eq:fw-one-emission-bspace-soft}
  \end{align}
  The ellipsis represents higher azimuthal harmonics and subleading isotropic terms.
  The finite first harmonic follows from the Bessel integral
  $\int_0^\infty (dk_T/k_T)J_1(bk_T)=1$. We incorporate it through the factor
  $1-2i[\alpha_s C_F/\pi]B_1\cos\phi_b$ multiplying the Sudakov exponential,
  following the electron--jet treatment~\cite{Hatta:2021jcd}.

  The acceptance dependence enters the Sudakov exponent through $B_0$, alongside the standard
  quark non-cusp contribution $-3/2$. With the running coupling, the exponent reads
  \begin{equation}
  S(b;\Delta\eta)
  =
  \int_{\mu_{b}}^{Q}\frac{d\mu}{\mu}
  \frac{\alpha_{s}(\mu)C_{F}}{\pi}
  \left[
  \ln\frac{Q^{2}}{\mu^{2}}
  +\ln\frac{Q^{2}}{P_{T}^{2}}
  -\frac{3}{2}
  +B_{0}(\Delta\eta)
  \right].
  \label{eq:fw-sudakov-Q}
  \end{equation}

  The common factor $e^{-S-S_{\rm NP}}$ combines global Sudakov resummation with
  the nonperturbative transverse dynamics specified in section~\ref{sec:numerics}.
  The azimuthally averaged and first-harmonic unpolarized structure functions are
  \begin{align}
  F_0(q_{T}) & =\frac{x_{B}}{2\pi}\sum_{q,\bar{q}}e_{q}^{2}\int_{0}^{\infty}db\,b\,J_{0}(bq_{T})\,e^{-S-S_{{\rm NP}}}\,\widetilde{f}_{1}^{\,q}(x_{B},b;\mu_{b}),
  \label{eq:fw-unpolarized-fuu}
  \\
  F_1(q_{T}) & =\frac{x_{B}}{2\pi}\sum_{q,\bar{q}}e_{q}^{2}\int_{0}^{\infty}db\,b\,J_{1}(bq_{T})\,e^{-S-S_{{\rm
  NP}}}\,\widetilde{f}_{1}^{\,q}(x_{B},b;\mu_{b})\frac{\alpha_{s}(\mu_{b})C_{F}}{\pi}B_{1}(\Delta\eta),
  \label{eq:fw-unpolarized-structure-functions}
  \end{align}
  Here $q$ labels flavor and $e_q$ is the corresponding electric charge.

  The Sivers and worm-gear modulations originate from the spin--momentum correlations in the quark TMDs.
  Their tensor structures in eq.~\eqref{eq:fw-bspace-correlators} give an extra power of $b$ and a $J_1$ weight,
  yielding
  \begin{align}
  F_{UT}(q_{T}) & =\frac{M_{p}x_{B}}{2\pi}\sum_{q,\bar{q}}e_{q}^{2}\int_{0}^{\infty}db\,b^{2}J_{1}(bq_{T})\,e^{-S-S_{{\rm
  NP}}}\,\widetilde{f}_{1T}^{\perp(1)q}(x_{B},b;\mu_{b}),
  \label{eq:fw-sivers-structure}
  \\
  F_{LT}(q_{T}) & =\frac{M_{p}x_{B}}{2\pi}\sum_{q,\bar{q}}e_{q}^{2}\int_{0}^{\infty}db\,b^{2}J_{1}(bq_{T})\,e^{-S-S_{{\rm
  NP}}}\,\widetilde{g}_{1T}^{(1)q}(x_{B},b;\mu_{b}).
  \label{eq:fw-spin-structure-functions}
  \end{align}
  Both spin-dependent structure functions share the same acceptance-dependent Sudakov factor
  as the unpolarized structure function as discussed earlier.

\subsection{Non-global evolution}
\label{sec:fw-ngl}

The Sudakov factor discussed above resums independent soft emissions. Radiation
correlated across the acceptance boundary gives an additional contribution to the
structure functions. A gluon emitted inside the window does not enter the recoil
directly, but acts as an extra color source that can radiate softer gluons outside.
Because the measurement assigns different weights to the two regions, a correlated
contribution survives real--virtual cancellation. These non-global logarithms (NGLs)
first appear at $\mathcal O(\alpha_s^2)$~\cite{Dasgupta:2001sh,Dasgupta:2002bw}.

In the large-$N_c$ limit, such an emission splits the dipole formed by the incoming
and struck quarks into two daughter dipoles. While the splitting conserves total
color charge, it modifies the angular distribution of softer radiation that resolves
the daughter dipoles. The BMS equation describes successive branchings, thereby
resumming these NGLs at leading-logarithmic accuracy~\cite{Banfi:2002hw}. Within
SCET, the color sources are represented by operators
containing multiple Wilson lines, whose renormalization-group evolution generates
the NGL resummation~\cite{Becher:2015hka,Becher:2016mmh}. This framework has been applied
to jet observables and to transverse-recoil resummation in boson--jet and dijet
production~\cite{Hatta:2009nd,Hatta:2013qj,Becher:2016omr,Becher:2017nof,Balsiger:2018ezi,Balsiger:2019tne,Chien:2019gyf,Fu:2026nkd}.

For the fiducial imbalance, transverse momenta outside the acceptance enter through
their vector sum, allowing cancellations between emissions. This differs from the
rapidity-dependent vetoes studied in SCET, which constrain scalar quantities such as
transverse energy or jet transverse momentum~\cite{Hornig:2017pud,Michel:2018hui}.
At $\mathcal O(\alpha_s^2)$, the azimuthal average and first harmonic needed for the
structure functions probe the same correlated emission kernel, but their different
recoil weights lead to distinct logarithmic behavior. For the fixed rapidity window
considered here, we derive the acceptance dependence of both projections analytically.
These fixed-order results provide the leading non-global contribution to the first
harmonic as well as an analytic reference for BMS resummation of the azimuthal
average~\cite{Banfi:2002hw}.

The recoil weights follow from the Fourier representation of the vector constraint,
using the same impact parameter $\boldsymbol b$ introduced earlier:
\begin{equation}
\delta^{(2)}\!\left(\boldsymbol{q}_{T}-\boldsymbol{\kappa}_{T}+\sum_{r\in\bar{\Omega}}\boldsymbol{k}_{rT}\right)=\int\frac{d^{2}\boldsymbol{b}}{(2\pi)^{2}}e^{i\boldsymbol{b}\cdot(\boldsymbol{q}_{T}-\boldsymbol{\kappa}_{T})}e^{i\boldsymbol{b}\cdot\sum_{r\in\bar{\Omega}}\boldsymbol{k}_{rT}}.\label{eq:fw-ngl-fourier-measurement}
\end{equation}
This representation assigns a nontrivial Fourier weight only to emissions in
$\bar{\Omega}$,
\begin{equation}
e^{i\boldsymbol{b}\cdot\sum_{r\in\bar{\Omega}}\boldsymbol{k}_{rT}}=\prod_{r}\left[\Theta_{\Omega}(k_{r})+\Theta_{\bar{\Omega}}(k_{r})e^{i\boldsymbol{b}\cdot\boldsymbol{k}_{rT}}\right]\label{eq:fw-ngl-exact-average}
\end{equation}

Consider two strongly ordered soft gluons with $P_T>k_{T1}>k_{T2}$, where the harder
gluon $k_1$ lies inside the acceptance and the softer gluon $k_2$ falls outside.
We measure their azimuths $\phi_1,\phi_2$ from the Born-quark direction, while
$\phi_b$ is measured from the scattered-electron transverse direction.
Taking $\boldsymbol\kappa_T=0$, the recoil becomes $\boldsymbol q_T=-\boldsymbol k_{2T}$.

The independent $C_F^2$ contribution is already included in the Sudakov factor.
The correlated term carries the color factor $C_FC_A/2$ in the antenna normalization
of eq.~\eqref{eq:fw-eikonal-antenna}. We replace $C_F\to C_A/2$ in the large-$N_c$ limit.
The change in the softer radiation pattern is then described by the two daughter
antennas minus the parent antenna. Using the dipole measure
$d\Pi_{ij}(k)=\frac{d\Omega_k}{4\pi}
\frac{1-\cos\theta_{ij}}{(1-\cos\theta_{ik})(1-\cos\theta_{kj})}$,
we define the angular coefficients for $n=0,1$ as
\begin{equation}
I_n(\Delta\eta)=
\int_{1\in\Omega}d\Pi_{BJ}(1)
\int_{2\in\bar{\Omega}}
\left[d\Pi_{B1}(2)+d\Pi_{1J}(2)-d\Pi_{BJ}(2)\right]
\cos(n\phi_2).
\label{eq:fw-ngl-angular-moments}
\end{equation}
Evaluating these angular integrals yields
\begin{equation}
\begin{aligned}
I_{0}(\Delta\eta) & =\frac{\pi^{2}}{6}-\frac{1}{2}\operatorname{Li}_{2}(1-e^{-\Delta\eta}),\\
I_{1}(\Delta\eta) & =\frac{1-3e^{-\Delta\eta}}{2e^{-\Delta\eta/2}}\ln(1-e^{-\Delta\eta})
 +\frac{\pi^{2}}{6}-\operatorname{Li}_{2}\!\left[\left(\frac{1-e^{-\Delta\eta/2}}{1+e^{-\Delta\eta/2}}\right)^{2}\right]\\
 & \quad+2e^{-\Delta\eta/2}\ln(2e^{-\Delta\eta/2})
 +2\operatorname{atanh}(e^{-\Delta\eta/2})\left[\ln\frac{4e^{-\Delta\eta/2}}{(1+e^{-\Delta\eta/2})^{2}}-1\right].
\end{aligned}
\label{eq:fw-ngl-angular-results}
\end{equation}
The detailed derivation is given in appendix~\ref{app:two-emission}.

The initial-dipole factor has the harmonic expansion
$\mathcal U^{(2)}=U_0^{(2)}-2i\,U_1^{(2)}\cos\phi_b+\cdots$.
Throughout this section, we use $\bar\alpha_s=C_A\alpha_s/\pi$.
Each emission coupling is evaluated at the corresponding transverse momentum.
We suppress this scale argument and the common dependence on $b,\Delta\eta$ below.
The recoil integrals then give
\begin{equation}
\begin{aligned}
U_0^{(2)}
&=1+I_0\int_0^{P_T}\frac{dk_{T1}}{k_{T1}}\bar{\alpha}_s
\int_0^{k_{T1}}\frac{dk_{T2}}{k_{T2}}\bar{\alpha}_s
[J_0(bk_{T2})-1],\\
U_1^{(2)}
&=I_1\int_0^{P_T}\frac{dk_{T1}}{k_{T1}}\bar{\alpha}_s
\int_0^{k_{T1}}\frac{dk_{T2}}{k_{T2}}\bar{\alpha}_s
J_1(bk_{T2}).
\end{aligned}
\label{eq:fw-ngl-running-two-emission}
\end{equation}

At fixed coupling, the harder-emission integral produces $\ln(P_T/k_{T2})$.
In the limit $bP_T\to\infty$, the two projections reduce to
\begin{equation}
\begin{aligned}
U_0^{(2)}
&\approx 1-\frac{\bar\alpha_s^2 I_0}{2}\ln^2\frac{P_T}{\mu_b},\\
U_1^{(2)}
&\approx \bar\alpha_s^2 I_1
\left(\ln\frac{P_T}{\mu_b}+2\ln2-1\right).
\end{aligned}
\label{eq:fw-ngl-numerical-expansion}
\end{equation}
The different logarithmic behavior of the two projections follows from their recoil
weights. For the azimuthal average, the virtual subtraction leaves a logarithm from
the softer emission after the oscillatory contribution averages out. Combined with
the harder-emission logarithm, this produces the double logarithm. By contrast, the
first harmonic has no virtual term; its $J_1$ integral converges at large $bk_{T2}$,
leaving only the single logarithm from the harder emission.

At recoil scales well below $P_T$, successive soft emissions generate enhanced
non-global logarithms in the azimuthal average. The double logarithm in $U_0^{(2)}$
is the first term in this series, which BMS evolution resums to all orders. At large $N_c$, each real emission inside the acceptance splits the parent dipole
into two daughters that subsequently evolve independently at lower transverse
momenta. Virtual corrections leave the parent dipole intact. We denote the dipole
distribution by $\mathcal G_{ij}(k_T,\boldsymbol b;\Delta\eta)$. Evolution ordered
in laboratory transverse momentum then takes the form
\begin{equation}
\frac{\partial\mathcal{G}_{ij}}{\partial\ln k_{T}}
=\bar\alpha_s\int_{\Omega}d\Pi_{ij}(k)\left[\mathcal{G}_{ik}\mathcal{G}_{kj}-\mathcal{G}_{ij}\right]
+\bar\alpha_s\int_{\bar{\Omega}}d\Pi_{ij}(k)\left[e^{i\boldsymbol{b}\cdot\boldsymbol{k}_{T}}-1\right]\mathcal{G}_{ij}.
\label{eq:fw-ngl-restricted-functional}
\end{equation}

Radiation outside the acceptance enters through the recoil weight and does not
initiate secondary branching. This contribution exponentiates as
$e^{-\mathcal R_{ij}}$, where
\begin{equation}
\mathcal{R}_{ij}(k_{T},\boldsymbol{b};\Delta\eta)
=\int^{k_{T}}\frac{dk_{T}'}{k_{T}'}\,\bar\alpha_s
\int_{\bar{\Omega}}d\Pi_{ij}(l)\left[1-e^{i\boldsymbol{b}\cdot\boldsymbol{k}_{T}'}\right].
\label{eq:fw-ngl-primary-radiator}
\end{equation}
Factoring out this contribution via $\mathcal G_{ij}=e^{-\mathcal R_{ij}}\mathcal U_{ij}$
leads to the BMS equation~\cite{Banfi:2002hw,Balsiger:2018ezi}:
\begin{equation}
\frac{\partial\mathcal{U}_{ij}}{\partial\ln k_{T}}
=\bar\alpha_s\int_{\Omega}d\Pi_{ij}(k)
\left[e^{-(\mathcal{R}_{ik}+\mathcal{R}_{kj}-\mathcal{R}_{ij})}
\mathcal{U}_{ik}\mathcal{U}_{kj}-\mathcal{U}_{ij}\right].
\label{eq:fw-bms-equation}
\end{equation}
The exponential factor is the ratio of the daughter and parent Sudakov factors.
At $b=0$, the recoil phase is unity and inclusivity enforces $\mathcal U_{ij}=1$.

For the azimuthal average $U_0$, we resum the non-global contribution using the
recoil weight $1-J_0(bk_T)$. We do not resum the non-global first harmonic.
Instead, we keep its leading $\mathcal O(\alpha_s^2)$ contribution,
$U_1=U_1^{(2)}$, from eq.~\eqref{eq:fw-ngl-running-two-emission}, with the full
$J_1$ weight and a coupling evaluated at each emission's transverse momentum.
The same equation gives the two-gluon term $U_0^{(2)}$, which we use as a
fixed-order reference for the numerical BMS evolution of $U_0$. The numerical
method is described in
appendix~\ref{app:numerical-recipe}.

The one-loop $\cos\phi$ term and the non-global contributions enter the factor
multiplying the Sudakov exponential. Through the first harmonic, this factor is
\begin{equation}
U_0-2i\left[\frac{C_F}{C_A}\bar\alpha_s(\mu_b)B_1U_0+U_1^{(2)}\right]
\cos\phi_b+\cdots.
\label{eq:fw-ngl-harmonic-matching}
\end{equation}

Including the large-$N_c$ NGL contribution, we obtain the structure functions used
in our phenomenology. Suppressing the common TMD arguments $(x_B,b;\mu_b)$, we have
\begin{align}
F_0 & =\frac{x_{B}}{2\pi}\sum_{q,\bar{q}}e_{q}^{2}\int_{0}^{\infty}db\,b\,J_{0}(bq_{T})e^{-S-S_{{\rm NP}}}\widetilde{f}_{1}^{\,q}\,U_{0},\label{eq:fw-ngl-F0}\\
F_1 & =\frac{x_{B}}{2\pi}\sum_{q,\bar{q}}e_{q}^{2}\int_{0}^{\infty}db\,b\,J_{1}(bq_{T})e^{-S-S_{{\rm NP}}}\widetilde{f}_{1}^{\,q}\left[\frac{C_F}{C_A}\bar\alpha_s(\mu_b)B_1U_0+U_1^{(2)}\right],\\
F_{UT} & =\frac{M_{p}x_{B}}{2\pi}\sum_{q,\bar{q}}e_{q}^{2}\int_{0}^{\infty}db\,b^{2}J_{1}(bq_{T})e^{-S-S_{{\rm NP}}}\widetilde{f}_{1T}^{\perp(1)q}U_{0},\\
F_{LT} & =\frac{M_{p}x_{B}}{2\pi}\sum_{q,\bar{q}}e_{q}^{2}\int_{0}^{\infty}db\,b^{2}J_{1}(bq_{T})e^{-S-S_{{\rm NP}}}\widetilde{g}_{1T}^{(1)q}U_{0}.
\label{eq:fw-ngl-matching}
\end{align}
For the spin modulations, we include the azimuthally averaged soft factor $U_0$.
The integrands of the polarized numerators contain an extra power of $b$ compared
with the unpolarized denominator. These different weights in the $b$ integrals
make the spin asymmetries sensitive to the common soft factors. In particular,
the polarized structure functions are more sensitive to large $b$, precisely
where the acceptance-dependent suppression is strongest.

\endgroup
\section{Phenomenology at HERA, the EIC and EicC}
\label{sec:phenomenology}

We present numerical predictions for the fiducial imbalance
at HERA, the EIC~\cite{Accardi:2012qut,AbdulKhalek:2021gbh}, and
EicC~\cite{Anderle:2021wcy,Xiao:2026tbs}. Varying the rapidity-window width,
we examine the unpolarized recoil spectrum and $\langle\cos\phi\rangle$, compare
predictions with and without NGLs, and study the Sivers and worm-gear spin
asymmetries at the EIC and EicC.

\subsection{Kinematics and TMD inputs}
\label{sec:numerics}

We study the three benchmark points listed in table~\ref{tab:born-benchmarks},
which span a broad range of hard scales and Bjorken $x$. The EicC point probes
lower $Q$ and larger $x_B$ than the EIC point. For all three colliders, we use
centered windows with $\Delta\eta=1,2,3$ .

\begin{table}[ht!]
\centering
\small
\begin{tabular}{lcccc}
\hline
 Collider & $(E_e,E_p)$ & $P_T$ & $Q$ & $x_B$  \\
\hline
 HERA & $(27.6,920)$ & $40.6$ & $78.8$ & $0.0832$\\
 EIC  & $(18,275)$   & $26.5$ & $51.4$ & $0.182$ \\
 EicC & $(3.5,20)$   & $5.15$ & $10$ & $0.485$ \\
\hline
\end{tabular}
\caption{Benchmark Born kinematics at $y=0.735$ and central quark pseudorapidity $\eta=0$. Beam energies, $P_T$, and $Q$ are given in $\mathrm{GeV}$.}
\label{tab:born-benchmarks}
\end{table}

The collinear parton distribution functions (PDFs) and next-to-leading-order
running coupling are taken from the central member of
\texttt{HERAPDF20\_NLO\_EIG}~\cite{H1:2015ubc}, with active quark and antiquark
flavors $u,d,s,c,b$. We denote this PDF input by $f_q$.

The large-$b$ behavior is regulated with the standard $b_\ast$ prescription~\cite{Collins:1984kg}, with
\begin{align}
 b_\ast=\frac{b}{\sqrt{1+b^2/b_{\max}^2}},
  b_{\max}=1.5\,\mathrm{GeV}^{-1},
 \mu_b=\frac{b_0}{b_\ast}.
 \label{eq:num-scale-profile}
\end{align}
In the numerical evaluation, the scale is restricted to $1\,\mathrm{GeV}\leq\mu_b\leq Q$. We evaluate the collinear PDFs and the coupling in the one-loop $B_1$ term at this scale and use it as the lower limit of the Sudakov integral in eq.~\eqref{eq:fw-sudakov-Q}. We use the physical impact parameter $b$ in the Hankel transforms and NGL evolution.

We model nonperturbative transverse dynamics at large $b$ with a common
exponent for all structure functions:
\begin{align}
 S_{\mathrm{NP}}(b,Q)
 &=g_q b^2+g_K\ln\frac{Q}{Q_0}\ln\frac{b}{b_\ast},
 &
 g_q&=0.106\,\mathrm{GeV}^2,
 \nonumber\\[-2pt]
 g_K&=0.42,
 & Q_0^2&=2.4\,\mathrm{GeV}^2.
 \label{eq:num-nonperturbative}
\end{align}
For a single incoming quark TMD, the parameters $g_q$ and $g_K$ are one half
of the Drell--Yan fit values in refs.~\cite{Su:2014wpa,Prokudin:2015ysa}.
We adopt this incoming-proton exponent
as the full nonperturbative transverse profile in all three channels.

For $1/b\gg\Lambda_{{\rm QCD}}$, the TMDs can be matched onto collinear
distributions.
At leading order, we use the collinear PDF at $\mu_b$ for the unpolarized TMD.
For the Sivers~\cite{Ji:2006ub,Ji:2006vf,Ji:2006br} and
worm-gear~\cite{Zhou:2009jm,Rein:2022odl} inputs, the small-$b$ relations for
each quark flavor are
\begin{align}
\widetilde{f}_{1T}^{\perp(1)}(x,b;\mu_{b})
&=-\frac{T_{F}(x,x;\mu_{b})}{2M_{p}},\label{eq:fw-tree-matching}\\
\widetilde{g}_{1T}^{(1)}(x,b;\mu_{b})
&=\tilde g(x;\mu_{b}).\label{eq:fw-g1t-boundary}
\end{align}
Here $T_F(x,x)$ is the diagonal Qiu--Sterman correlation
function~\cite{Qiu:1991pp,Qiu:1998ia}, while $\tilde g(x)$ is the chiral-even
twist-three function associated with the worm-gear moment~\cite{Zhou:2009jm}
and enters collinear twist-three calculations of $A_{LT}$~\cite{Liang:2012rb,Metz:2012fq}.
The sign in the Sivers relation corresponds to the future-pointing SIDIS gauge
link; the naive-$T$-even worm-gear function has no corresponding sign reversal.

The evolution of $T_F$ couples it to off-diagonal quark--gluon and three-gluon
correlators~\cite{Kang:2008ey,Vogelsang:2009pj,Braun:2009mi,Schafer:2012ra,Kang:2012em,Ma:2012xn,Kang:2012ns,Dai:2014ala,Yoshida:2016tfh}.
Similarly, the evolution of $\tilde g$ involves other quark--gluon
correlations~\cite{Zhou:2008mz}. Phenomenological inputs for these additional
correlations are lacking. We therefore
approximate the scale dependence of $T_F$ and $\tilde g$ by that of the
corresponding unpolarized PDF under DGLAP evolution. With this prescription, we multiply the fitted flavor-dependent $x$ prefactors of
refs.~\cite{Echevarria:2020hpy,Bhattacharya:2021twu} by the central
\texttt{HERAPDF20} distribution at $\mu_b$.

For the Sivers input, we use the initial-scale flavor-dependent $x$ shape from Fit~1 of
ref.~\cite{Echevarria:2020hpy} and write the Qiu--Sterman function at $\mu_b$ as
\begin{align}
 T_F^q(x,x;\mu_b)
 &=\mathcal N_q(x)\,f_q(x;\mu_b),
 \nonumber\\
 \mathcal N_q(x)
 &=N_q\,
   \frac{(\alpha_q+\beta)^{\alpha_q+\beta}}
        {\alpha_q^{\alpha_q}\beta^\beta}
   x^{\alpha_q}(1-x)^\beta,
 \qquad \beta=5.13.
 \label{eq:num-sivers-input}
\end{align}
The central light-quark and antiquark parameters are listed in
table~\ref{tab:num-sivers-parameters}, while charm and bottom contributions are
set to zero.

\begin{table}[h!]
 \centering
 \small
 \begin{tabular}{c|rrrrrr}
  \hline
  flavor & $u$ & $d$ & $s$ & $\bar u$ & $\bar d$ & $\bar s$ \\
  \hline
  $N_q\,[\mathrm{GeV}]$ & $0.077$ & $-0.152$ & $0.167$ & $-0.033$ & $-0.069$ & $-0.002$ \\
  $\alpha_q$             & $0.967$ & $1.19$   & $0.936$ & $0.936$  & $0.936$  & $0.936$ \\
  \hline
 \end{tabular}
 \caption{Central parameters for the Sivers input from Fit~1 in Table~2 of ref.~\cite{Echevarria:2020hpy}, with common exponent $\beta=5.13$.}
 \label{tab:num-sivers-parameters}
\end{table}

For the worm-gear input, we follow the $b$-space treatment of
ref.~\cite{Kang:2021ffh} and use the flavor-dependent $x$ shape from the
central weighted-$\chi^2$ fit of ref.~\cite{Bhattacharya:2021twu}, writing
its first moment at $\mu_b$ as
\begin{align}
 g_{1T}^{(1)q}(x;\mu_b)
 &=\mathcal G_q(x)\,f_q(x;\mu_b),
 &
 \mathcal G_q(x)&=C_q x^{1.9}(1-x).
 \label{eq:num-g1t-input}
\end{align}
Here $C_u=1.32$, $C_d=-1.64$, and $C_q=0$ for all other flavors.
The coefficients are obtained from the ref.~\cite{Bhattacharya:2021twu}.

The acceptance dependence comes from radiation outside the window. For the
non-global contribution, we combine all-order large-$N_c$ evolution for $U_0$
with the leading first-harmonic contribution $U_1^{(2)}$, as specified in
eq.~\eqref{eq:fw-ngl-harmonic-matching}. We evaluate these factors with the
same running coupling as above, frozen below $1\,\mathrm{GeV}$;
appendix~\ref{app:numerical-recipe} presents the numerical BMS implementation. To isolate
the effect of NGLs, we compare with a reference prediction in which $U_0=1$
and $U_1^{(2)}=0$, including the one-loop $B_1$ term and all other inputs.

\subsection{Acceptance dependence}
\label{sec:results}

\begin{figure}[ht!]
  \centering
  \includegraphics[width=0.485\textwidth]{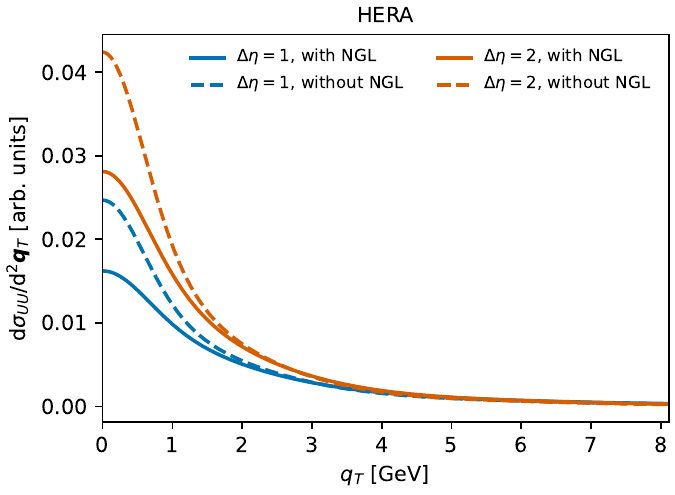}\hfill
  \includegraphics[width=0.485\textwidth]{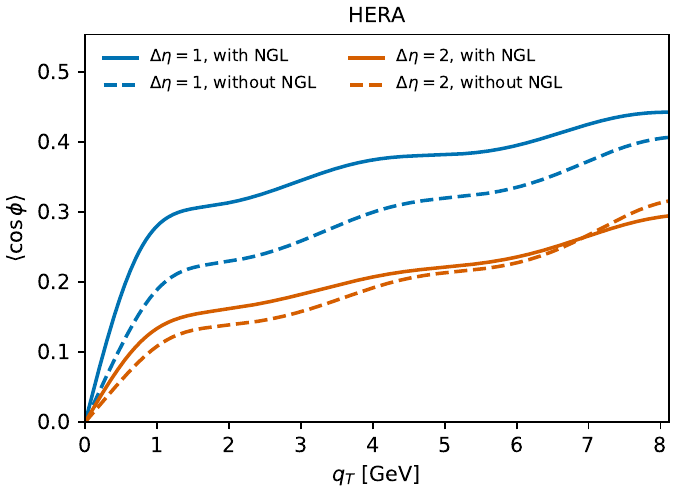}
  \caption{HERA predictions as functions of transverse recoil $q_T$ for centered acceptance windows with $\Delta\eta=1,2$. Left: azimuthally averaged unpolarized cross section, shown with the common normalization suppressed. Right: unpolarized first azimuthal harmonic $\langle\cos\phi\rangle$. Solid and dashed curves show predictions with and without NGLs, respectively, using the same remaining inputs.}
  \label{fig:hera-observables}
\end{figure}

Figure~\ref{fig:hera-observables} shows the HERA recoil spectrum and first
azimuthal moment for two window widths. A wider window includes more
radiation near the struck quark in the hadronic sum, reducing the measured
soft recoil. The spectrum consequently shifts toward smaller $q_T$, with more
events at low recoil and fewer at large recoil. This is a redistribution of the
same DIS sample: we change only the measured $q_T$, without a radiation veto.

The reduced soft recoil also weakens the $\cos\phi$ modulation. At HERA,
$\langle\cos\phi\rangle$ vanishes at $q_T=0$ by rotational symmetry and increases
over the displayed range, with smaller values for the wider window throughout.
For all three colliders, the moment at $q_T=1\,\mathrm{GeV}$ decreases by about
a factor of two as $\Delta\eta$ increases from $1$ to $2$, and decreases further
at $\Delta\eta=3$ (table~\ref{tab:results-q1-summary}). The one-loop coefficient
$B_1(\Delta\eta)$ shows the same trend, decreasing exponentially at large
window widths.

At fixed window width, NGLs suppress the unpolarized spectrum at low $q_T$
and shift it toward larger recoil. A real gluon inside the window does not
contribute directly to the measured recoil, but can emit a softer gluon outside
it. Such emissions are enhanced when the two gluons are nearly collinear on
opposite sides of the acceptance boundary. The radiation pattern changes,
but the total color charge is conserved. Although the radiation is not enhanced
in every direction, the integrated change relative to the parent dipole is
positive, $I_0(\Delta\eta)>0$. Combined with the real--virtual recoil weight
$J_0(bk_{T2})-1\leq0$ in eq.~\eqref{eq:fw-ngl-running-two-emission}, this gives
a negative two-loop correction to $U_0$, consistent with the low-recoil
suppression.

The NGL correction to $\langle\cos\phi\rangle=F_1/F_0$ depends on the relative
changes in the two structure functions. At HERA, NGLs enhance the moment
throughout the displayed range for $\Delta\eta=1$, and at small recoil for
$\Delta\eta=2$.

\begin{table}[ht!]
 \centering
 \small
 \begin{tabular}{lrrr}
  \hline
$\langle\cos\phi\rangle$ & $\Delta\eta=1$ & $\Delta\eta=2$ & $\Delta\eta=3$ \\
  \hline
  HERA 
    & $0.2799$ & $0.1337$ & $0.0740$ \\
 EIC
    & $0.3422$ & $0.1632$ & $0.0905$ \\
  EicC 
    & $0.5815$ & $0.2883$ & $0.1651$  \\
  \hline
 \end{tabular}
  \caption{$\langle\cos\phi\rangle$ at $q_T=1\,\mathrm{GeV}$ for the HERA, EIC, and EicC benchmarks with centered pseudorapidity windows $\Delta\eta=1,2,3$. The predictions include global Sudakov resummation and NGLs.}
 \label{tab:results-q1-summary}
\end{table}

\begin{figure}[ht!]
  \centering
  \includegraphics[width=0.485\textwidth]{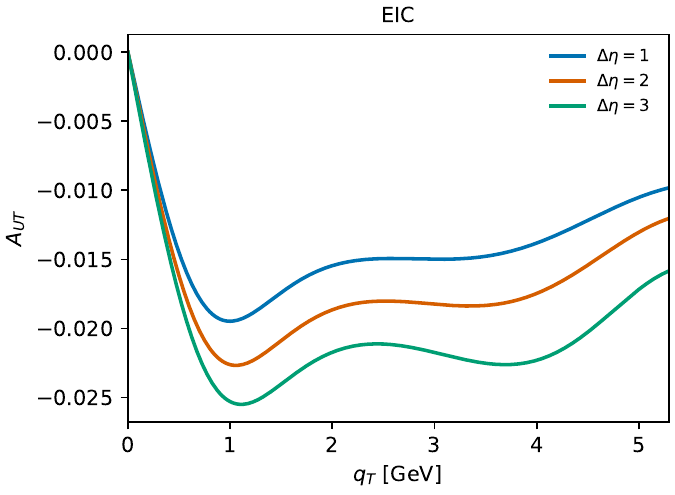}\hfill
  \includegraphics[width=0.485\textwidth]{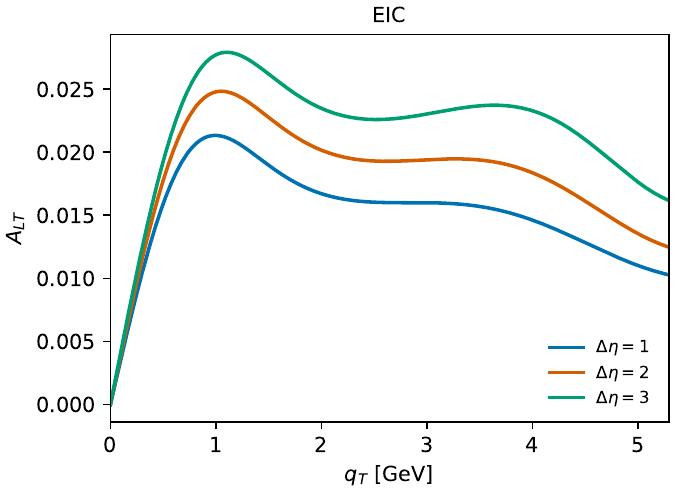}
  \caption{EIC spin asymmetries versus $q_T$ for centered windows with $\Delta\eta=1,2,3$, including global Sudakov resummation and NGLs. Left: Sivers single-spin asymmetry $A_{UT}$ in $\sin(\phi-\phi_S)$. Right: longitudinal--transverse double-spin asymmetry $A_{LT}$ in $\cos(\phi-\phi_S)$.}
  \label{fig:eic-spin-asymmetries}
\end{figure}

\begin{figure}[ht!]
  \centering
  \includegraphics[width=0.485\textwidth]{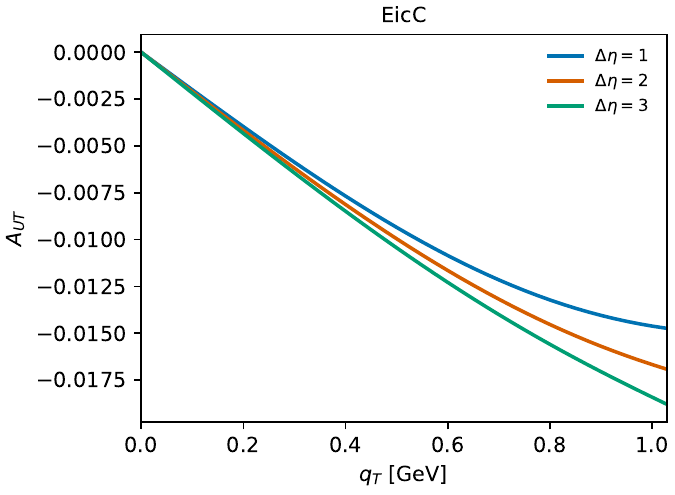}\hfill
  \includegraphics[width=0.485\textwidth]{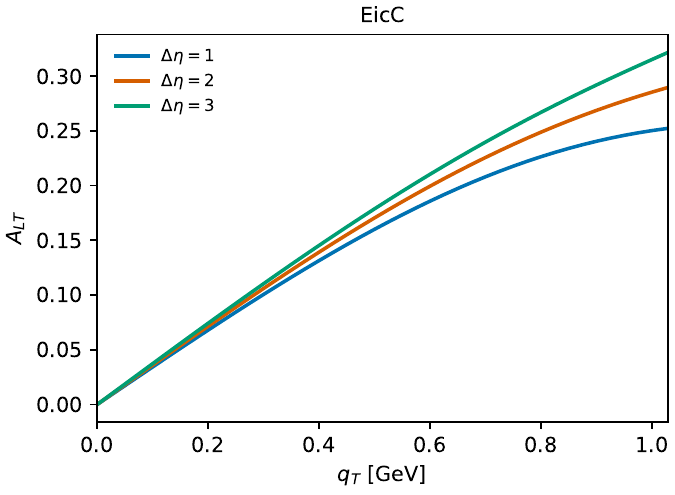}
  \caption{EicC spin asymmetries versus $q_T$ for centered windows with $\Delta\eta=1,2,3$, including global Sudakov resummation and NGLs. Left: Sivers single-spin asymmetry $A_{UT}$ in $\sin(\phi-\phi_S)$. Right: longitudinal--transverse double-spin asymmetry $A_{LT}$ in $\cos(\phi-\phi_S)$.}
  \label{fig:eicc-observables}
\end{figure}

The spin asymmetries have the opposite acceptance dependence: at fixed recoil,
both $|A_{UT}|$ and $|A_{LT}|$ increase with $\Delta\eta$ for the EIC and EicC
benchmarks (figures~\ref{fig:eic-spin-asymmetries} and~\ref{fig:eicc-observables}).
Whereas soft radiation generates the unpolarized $\cos\phi$ modulation,
azimuthally symmetric soft recoil dilutes the spin modulations arising from
the intrinsic correlations described by $f_{1T}^\perp$ and $g_{1T}$. Widening
the window reduces this dilution. Combined with the redistribution of events
toward lower recoil, this gives both more events and larger spin asymmetries
in the region most sensitive to nonperturbative proton structure.

At the EIC benchmark, $A_{UT}$ is negative and $A_{LT}$ positive, both peaking
in magnitude around $q_T\sim 1$--$1.5\,\mathrm{GeV}$. The signs follow from the
phenomenological inputs: the Sivers function has the sign structure extracted
from SIDIS, and the charge-weighted $g_{1T}$ input is positive, as discussed in
section~\ref{sec:numerics}.

The EicC benchmark probes lower $Q$ and larger $x_B$ and yields a larger
worm-gear asymmetry with the inputs used here. For $\Delta\eta=1$, for example,
$A_{LT}$ at $q_T=1\,\mathrm{GeV}$ is about $0.25$ at EicC, compared with about
$0.02$ at the EIC. The relative sizes of the Sivers and worm-gear asymmetries
reflect, in part, their different $x$ dependences: the $g_{1T}$ input contains
the factor $x^{1.9}(1-x)$, while the Sivers input has the stronger large-$x$
suppression $(1-x)^{5.13}$. Flavor weights and evolution intervals also
contribute to the differences between colliders. These predictions illustrate
how EicC measurements at larger $x$ can complement those at the EIC.

\begin{samepage}
Measuring the charges of final-state may provide additional sensitivity to quark flavor and help disentangle their contributions
to the spin asymmetries. We leave a
quantitative study to future work.

\end{samepage}

\section{Conclusions}
\label{sec:conclusions}

We have developed a framework for studying TMD observables through the fiducial imbalance in DIS,
with predictions for HERA, EIC, and EicC kinematics. By varying the rapidity window, this observable
controls radiative recoil without jet reconstruction or an explicit radiation veto.

The framework combines TMD factorization and global Sudakov resummation with a systematic treatment
of NGLs. The azimuthally averaged NGL contribution is resummed to all orders at leading-logarithmic
accuracy in the large-$N_c$ limit using BMS evolution, while the leading NGL correction to the first
azimuthal harmonic is included at $\mathcal{O}(\alpha_s^2)$. We have derived analytic expressions for
the acceptance-dependent coefficients $B_0(\Delta\eta)$ and $B_1(\Delta\eta)$, which enter the global
Sudakov exponent and one-loop harmonic matching, respectively, as well as for the NGL angular
coefficients $I_0$ and $I_1$. These results determine the fixed-order acceptance dependence and
provide a reference for the numerical BMS evolution.

For the kinematics studied, widening the window increases both the magnitudes of the Sivers and
worm-gear asymmetries and the event yield at low $q_T$, where sensitivity to nonperturbative
structure is greatest. Crucially, these gains arise from reduced soft-recoil dilution and a
redistribution of the same DIS event sample, rather than a radiation veto. At the EIC benchmark,
both asymmetries peak in magnitude around $q_T\sim 1$--$1.5\,\mathrm{GeV}$ and increase monotonically
in magnitude with $\Delta\eta$ over the range studied. At EicC, the larger Bjorken $x$ and lower $Q$
probe the valence region, where our predictions yield a particularly pronounced worm-gear asymmetry,
offering a promising complement to EIC measurements.

The unpolarized moment $\langle\cos\phi\rangle$ is sensitive to non-global effects
with reduced common normalization uncertainties. At $q_T=1\,\mathrm{GeV}$, it decreases by approximately
a factor of two as $\Delta\eta$ increases from $1$ to $2$ for all three collider benchmarks, and
decreases further at $\Delta\eta=3$. In the HERA predictions, NGL corrections enhance the moment at
small recoil for the narrow windows considered. Measuring this acceptance dependence within a single
dataset would provide a stringent test of the combined global and non-global contributions.

The framework we developed also provides a basis for extensions to heavy-flavor production,
photon-tagged events, and forward particle production. The fiducial imbalance thus offers a powerful complement to SIDIS and electron--jet
measurements, opening a practical route to the three-dimensional momentum and spin structure of
the nucleon.

\acknowledgments

This work was supported by the National Natural Science Foundation of China under Grant
Nos.~12175118, 12321005, and 12575091 (J.Z.) and by the Postdoctoral Innovation Program
of Shandong Province under Project No.~SDCX-ZG-202603015 (S.L.).

\appendix

\section{Angular integrals for two soft gluons}
\label{app:two-emission}

The coefficients $I_0(\Delta\eta)$ and $I_1(\Delta\eta)$ of
eq.~\eqref{eq:fw-ngl-angular-moments} arise from the correlated emission of two strongly ordered
soft gluons, the harder one lying inside the fiducial region, $1\in\Omega$, and the softer one
outside it, $2\in\bar{\Omega}$. The soft-radiation contribution to the recoil comes from $k_2$
alone, but its angular distribution is correlated with $k_1$: it is this correlation between
their directions that these coefficients encode, weighted by unity for $I_0$ and by
$\cos\phi_2$ for $I_1$.

For the angular integrations it is convenient to replace the angular coordinates
$(\eta_a,\phi_a)$ by the complex variable $z_a=e^{-\eta_a+i\phi_a}$. In this stereographic
projection rapidity fixes the distance from the origin, $|z_a|=e^{-\eta_a}$, while the
azimuth is unchanged. Infinite positive rapidity maps to the origin, whence the incoming-beam
direction is $z_B=0$; with the Born quark at central rapidity and azimuths measured from its
direction, $z_J=1$. The fiducial window $|\eta|<\Delta\eta/2$ is now the annulus
$\Omega:e^{-\Delta\eta/2}<|z|<e^{\Delta\eta/2}$, whose complement splits into an inner disk
$|z|<e^{-\Delta\eta/2}$ and an exterior region $|z|>e^{\Delta\eta/2}$ that may be treated
separately. Emission at $z$ off a dipole with endpoints $z_i$, $z_j$ then carries the angular
measure
\begin{equation}
d\Pi_{ij}(z)=\frac{d^{2}z}{2\pi}\frac{|z_{i}-z_{j}|^{2}}{|z_{i}-z|^{2}|z-z_{j}|^{2}}.\label{eq:fw-ngl-stereographic-measure}
\end{equation}

Consider first $I_0$, with $z_1$ held fixed. If both dipole endpoints lie outside the inner
disk, $|z_i|,|z_j|>e^{-\Delta\eta/2}$, the integral over that region is~\cite{Hatta:2009nd}
\begin{equation}
\int_{|z_{2}|<e^{-\Delta\eta/2}}d\Pi_{ij}(z_{2})=\frac{1}{2}\ln\frac{|z_{i}\bar{z}_{j}-e^{-\Delta\eta}|^{2}}{(|z_{i}|^{2}-e^{-\Delta\eta})(|z_{j}|^{2}-e^{-\Delta\eta})},\label{eq:fw-ngl-disk-integral}
\end{equation}
with a bar denoting complex conjugation. This covers the $1J$ dipole at once, since
$z_1\in\Omega$ and $z_J=1$ both lie outside the inner disk. For the $B1$ and $BJ$ terms it does
not apply directly: $z_B=0$ falls inside the integration domain, and the two disk integrals
carry the same beam-collinear divergence. Only their difference is finite, so we combine
the integrands before integrating. The exterior region $|z_2|>e^{\Delta\eta/2}$ is brought
back to the inner disk by the inversion $z\mapsto1/\bar z$.

Putting $z_1=re^{i\phi_1}$ and adding the two regions gives
\begin{equation}
\begin{aligned}
&I_{0}(\Delta\eta)={}\int_{e^{-\Delta\eta/2}}^{e^{\Delta\eta/2}}\frac{dr}{r}
\int_{0}^{2\pi}\frac{d\phi_{1}}{2\pi}\frac{1}{1-2r\cos\phi_{1}+r^{2}}\\
&\quad\times\Bigg[\ln\left|1-\frac{e^{-\Delta\eta}}{r}e^{i\phi_{1}}\right|
+\ln\left|1-e^{-\Delta\eta}re^{i\phi_{1}}\right|
-\ln\left(1-\frac{e^{-\Delta\eta}}{r^{2}}\right)
-\ln(1-e^{-\Delta\eta}r^{2})\Bigg].
\end{aligned}
\end{equation}
After splitting the radial integral at $r=1$ and mapping $1<r<e^{\Delta\eta/2}$ onto
$e^{-\Delta\eta/2}<r<1$ by $r\to1/r$, we carry out the azimuthal integrations on the resulting
interval with the Fourier series
\begin{equation}
\begin{aligned}\frac{1}{1-2r\cos\phi_{1}+r^{2}} & =\frac{1+2\sum_{m\geq1}r^{m}\cos(m\phi_{1})}{1-r^{2}},\\
\ln|1-re^{i\phi_{1}}| & =-\sum_{m\geq1}\frac{r^{m}}{m}\cos(m\phi_{1}),\qquad0<r<1.
\end{aligned}
\label{eq:fw-ngl-angular-series}
\end{equation}
Only the diagonal products of modes survive the $\phi_1$ integration, leaving the
one-dimensional integral
\begin{align}
I_{0}(\Delta\eta) & =\int_{e^{-\Delta\eta/2}}^{1}\frac{dr}{1-r^{2}}\left(\frac{1}{r}+r\right)\ln\frac{1-e^{-\Delta\eta}}{1-e^{-\Delta\eta}/r^{2}}.\label{eq:fw-ngl-isotropic-radial}
\end{align}
Its $1/r$ term descends from the segment $e^{-\Delta\eta/2}<r<1$, the $r$ term from the
inversion of $1<r<e^{\Delta\eta/2}$. Under the substitution
$\xi=(1-e^{-\Delta\eta}/r^2)/(1-e^{-\Delta\eta})$ the two evaluate to
$\pi^2/12$ and $\pi^2/12-\tfrac{1}{2}\operatorname{Li}_2(1-e^{-\Delta\eta})$, whose sum is the
$I_0(\Delta\eta)$ of eq.~\eqref{eq:fw-ngl-angular-results}.

For the first harmonic we integrate over the harder-gluon direction first, leaving
$\cos\phi_2$ as a weight in what remains. The correlated angular kernel is symmetric under
interchange of the two emission directions, so
\begin{equation}
\begin{aligned}I_{1}={} & \int_{1\in\Omega}d\Pi_{BJ}(1)\int_{2\in\bar{\Omega}}\Big[d\Pi_{B1}(2)+d\Pi_{1J}(2)-d\Pi_{BJ}(2)\Big]\cos\phi_{2}\\
={} & \int_{2\in\bar{\Omega}}d\Pi_{BJ}(2)\cos\phi_{2}\int_{1\in\Omega}\Big[d\Pi_{B2}(1)+d\Pi_{2J}(1)-d\Pi_{BJ}(1)\Big].
\end{aligned}
\label{eq:fw-ngl-first-reordered}
\end{equation}
Nothing changes in this step but the order of the angular integrations; the
transverse-momentum ordering and the assignment of the two emissions to $\Omega$ and
$\bar\Omega$ are as before. We expand the angular factors in Fourier modes once more,
keeping the terms that survive the $\cos\phi_2$ projection. The radial contributions related
by $r\leftrightarrow1/r$ then combine into
\begin{equation}
\begin{aligned}I_{1}(\Delta\eta)=\int_{e^{-\Delta\eta/2}}^{1}dr\,\Bigg[ & \frac{2(1+r^{2})}{r(1-r^{2})}\left\{ r\operatorname{atanh}\frac{e^{-\Delta\eta/2}}{r}-\operatorname{atanh}(e^{-\Delta\eta/2})\right\} \\
 & -\frac{1+r^{2}}{r^{2}}\operatorname{atanh}(re^{-\Delta\eta/2})+\frac{e^{-\Delta\eta/2}}{r}\Bigg].
\end{aligned}
\label{eq:fw-ngl-first-radial}
\end{equation}
The dilogarithmic piece is reduced by the substitution
$\xi=[(1+e^{-\Delta\eta/2})/(1-e^{-\Delta\eta/2})](1-r)/(1+r)$:
\begin{equation}
\begin{aligned} & 4\int_{e^{-\Delta\eta/2}}^{1}\frac{dr}{1-r^{2}}\left[\operatorname{atanh}\frac{e^{-\Delta\eta/2}}{r}-\operatorname{atanh}(e^{-\Delta\eta/2})\right]\\
 & \quad=\int_{0}^{1}\frac{d\xi}{\xi}\left\{ \ln\left[1-\left(\frac{1-e^{-\Delta\eta/2}}{1+e^{-\Delta\eta/2}}\right)^{2}\xi\right]-\ln(1-\xi)\right\} \\
 & \quad=\frac{\pi^{2}}{6}-\operatorname{Li}_{2}\left[\left(\frac{1-e^{-\Delta\eta/2}}{1+e^{-\Delta\eta/2}}\right)^{2}\right].
\end{aligned}
\label{eq:fw-ngl-first-dilogarithm}
\end{equation}
Adding to this the remaining terms of eq.~\eqref{eq:fw-ngl-first-radial} yields
$I_1(\Delta\eta)$ as quoted in eq.~\eqref{eq:fw-ngl-angular-results}.

\section{Numerical implementation of the BMS evolution}
\label{app:numerical-recipe}

$U_0$ is evaluated with a large-$N_c$ soft-gluon dipole shower, in the Monte Carlo formulation
of non-global evolution~\cite{Dasgupta:2001sh,Banfi:2002hw,Balsiger:2018ezi}. For the azimuthally averaged component $U_0$, the recoil measurement gives the weight
$1-J_0(bk_T)$.Emissions inside the acceptance are generated as
successive dipole splittings; radiation outside it enters only through a weight carried by each
branching history.

The angular integrations use the dipole measure of appendix~\ref{app:two-emission}, in the
stereographic coordinates of ref.~\cite{Schwartz:2014wha}, with $z_B=0$ and $z_J=1$.
Emissions are generated in $\Omega$ only, with a disk of dimensionless radius $0.01$ around
each endpoint of the emitting dipole removed, and the same endpoint cutoff is applied to
the real and virtual terms.

Each history starts from the Born dipole $(B,J)$ at $k_T=P_T$ and evolves towards lower
transverse momenta in the variable
$t=\int_{k_T}^{P_T}(dk/k)\bar\alpha_s$, with $\bar\alpha_s=C_A\alpha_s/\pi$. The coupling runs
with the transverse momentum appearing in the integral and is frozen below $1\,\mathrm{GeV}$.
Once the coupling has been absorbed into $t$, the differential branching probability is
$dt\,d\Pi_{ij}(z)$. A configuration of dipoles at $t_1$ evolves to $t_2>t_1$ with no emission in
$\Omega$ with probability
\begin{equation}
  \exp\left[-(t_2-t_1)\sum_{(i,j)}\int_{\Omega}d\Pi_{ij}\right],
  \label{eq:bms-no-emission}
\end{equation}
the sum running over the active dipoles and each angular integral taken with the cutoff just
described. The next emission, from dipole $(i,j)$ in direction $z$ at $t_2$, therefore occurs
with differential probability
\begin{equation}
  \exp\left[-(t_2-t_1)\sum_{(k,l)}\int_{\Omega}d\Pi_{kl}\right]\,
  dt_2\,d\Pi_{ij}(z),
  \label{eq:bms-next-emission}
\end{equation}
the exponential accounting for the absence of earlier radiation in $\Omega$ and
$dt_2\,d\Pi_{ij}(z)$ specifying the branching itself; this distribution is sampled with the
Sudakov veto algorithm~\cite{Platzer:2011dq}. Every emission replaces the parent dipole
$(i,j)$ by the daughters $(i,z)$ and $(z,j)$, which then evolve independently at lower scales.
Iterating down to the infrared cutoff generates one ordered branching history.

Between successive emissions the set of active dipoles is fixed, so their Sudakov exponents for
radiation into $\bar\Omega$ simply add. Writing $V_{ij}=\int_{\bar\Omega}d\Pi_{ij}$ for the
angular integral of a single dipole, and dividing out the azimuthally averaged Born-dipole
Sudakov factor associated with eq.~\eqref{eq:fw-ngl-primary-radiator}, we obtain $U_0$ as an
average over branching histories,
\begin{equation}
  U_0(b;\Delta\eta)\simeq
  \left\langle\exp\left\{-\int_{k_{T,\min}}^{P_T}\frac{dk_T}{k_T}
  \bar\alpha_s[1-J_0(bk_T)]
  \left[\sum_{(i,j)}V_{ij}-V_{BJ}\right]\right\}\right\rangle.
  \label{eq:bms-monte-carlo-average}
\end{equation}
Here $b$ is the usual impact parameter, the sum runs over the dipoles present at scale
$k_T$ along a given history, and the brackets denote the Monte Carlo average. The angular
integrals for the dipole containing the beam direction and for the Born dipole share the same
beam-collinear singularity, which cancels in their difference. The angular integrals over
$\bar\Omega$ are done analytically, using eq.~\eqref{eq:fw-ngl-disk-integral} together with the
inversion of the exterior region. For a Born quark at central rapidity and the symmetric
acceptance used here the radiator difference is nonnegative, so with $1-J_0(bk_T)\ge0$ every
weight lies between zero and one and all weights reach unity at $b=0$. Since $b$ enters only through the weights, the same branching histories serve at every value of
$b$, and observable ratios are formed from the mean numerator and denominator.

\bibliographystyle{JHEP}
\bibliography{ref}

\end{document}